\documentclass[10pt]{article}

\usepackage{authblk}

\usepackage{graphicx} 
\usepackage[scriptsize,nooneline,hang]{caption}
\usepackage[hang,nooneline,scriptsize]{subfigure}
\usepackage{pstricks,pstricks-add}
\usepackage{pst-plot}
\usepackage{color}
\usepackage{epstopdf}
\usepackage{float}

\usepackage{amsmath,amsfonts,bbm,dsfont,mathrsfs} 

\usepackage{hyperref}

\newcommand{\tr}{\tilde{r}}
\newcommand{\dd}{{\mathrm{d}}}
\newcommand{\ta}{\tilde{a}}
\newcommand{\tQ}{\tilde{Q}}
\newcommand{\tL}{\tilde{L}}
\newcommand{\tLambda}{\tilde{\Lambda}}

\begin{document}
\title{Analytic solutions of the geodesic equation for Reissner-Nordstr\"{o}m-(anti-)de Sitter black holes surrounded by different kinds of regular and exotic matter fields} 

\author{Arindam Kumar Chatterjee, Kai Flathmann, Hemwati Nandan and Anik Rudra \footnote{Note that the authors are presented in alphabetical order.}}
\affil{Department of Physics, West Bengal State University, \\Barasat, Kolkata-700126, India}
\affil{Institut f\"ur Physik, Universit\"at Oldenburg, D--26111 Oldenburg, Germany}
\affil{Department of Physics, Gurukul Kangri Vishwavidyalaya, \\ Haridwar 249407, India}
\affil{Department of Physics, H. N. B. Garhwal University, S.R.T. Campus, Uttarakhand-249199, India}

\maketitle
\begin{abstract}
The purpose of this study is the derivation of the equation of motion for particles and light in the spacetime of Reissner-Nordstr\"{o}m-(anti-)de Sitter black holes surrounded by different kinds of regular and exotic matter fields. The complete analytical solutions of the geodesic equations are given in terms of the elliptic Weierstra{\ss} $\wp$-function and the hyperelliptic Kleinian $\sigma$-function. Finally after analyzing the geodesic motion of test particles and light using parametric diagrams and effective potentials, we present a list of all possible orbits.
\end{abstract}

\section{Introduction}
Recent high precession astronomical observations confirm the fact that the Universe is expanding at an accelerating rate. The observations also predict the existence of some form of energy with a large negative pressure called dark energy \cite{Perlmutter:1999,Riess:1998,Spergel:2007,Tegmark:2004,Seljak:2005}. Dark energy is supposed to be the origin behind the accelerated expansion of the Universe at a global scale. There are two proposed forms of dark energy. The first and simplest candidate is the cosmological constant \cite{Padmanabhan:2003}. A different possibility is dynamical scalar field models such as quintessence \cite{Carroll:1998}, chameleon \cite{Khoury:2004}, K-essence \cite{Armendariz:2000}, tachyon \cite{Padmanabhan:2002}, phantom \cite{Caldwell:2002}, and dilaton \cite{Gasperini:2001}. In an astrophysical scenario, this kind of dark energy should cause gravitational effects (e.g deflection of light coming from a distance star) and should therefore be taken into consideration in this context. The study of the geodesic motion is one way to understand the physical effects of the gravitational field and the influence of dark matter on the geometry of the spacetime.\\
Although the majority of gravitational effects can be discussed using approximations and numerical methods, a robust and systematic study of all effects can only be achieved by using analytical methods. The motion of test particles provides an experimentally feasible way to study the gravitational fields of black holes. Predictions about observables (such as light deflection, Shapiro time delay and perihelion shift) can be studied and compared with observations. In 1931 Hagihara \cite{Hagihara:1931} solved the geodesic equation for the Schwarzschild spacetime using the Weierstrass elliptic function. This analytical tool was further advanced by the application of hyperelliptic functions when higher order polynomials $(>4)$ occur in the geodesic equation. Here functions are based on the solution of the Jacobi inversion problem \cite{Hackmann:2008,Hackmann1:2008}. There  are many works dedicated to the literature of analytic study in the geodesic motion. It is not possible to include due to the large volume of existing work but we mention a few of them we found interesting. In four dimensions it was applied to the Schwarzschild-de Sitter \cite{Hackmann:2010}, Kerr-deSitter spacetime \cite{Hackmann2:2008} and in higher dimensions to Schwarzschild, Schwarzschild-anti-de Sitter, Reissner-Nordstr\"{o}m, Reissner-Nordstr\"{o}m-anti-de Sitter \cite{Hackmann:2009} and Myers-Perry spacetimes \cite{Enolski:2011,Kagramanova:2012}. Moreover this analytical method is used for some spacetime such as $f(R)$ gravity \cite{Soroushfar:2015}, GMGHS black holes \cite{Soroushfar:2016a}, BTZ \cite{Soroushfar:2016b}, Static cylindrically symmetric conformal spacetime \cite{Hoseini:2016}, Kerr-Newman-(A)dS spacetime and the rotating charged black hole spacetime in $f(R)$ gravity \cite{Soroushfar:2016c}, Einstein-Maxwell-dilaton-axion black holes \cite{Flathmann:2015} and $U(1)^2$ dyonic rotating black holes \cite{Flathmann:2016}.\\
Though the effect of dark energy is negligible at our local Universe, its existence cannot be completely ignored for various objects like black holes in our Universe at any scale. It is noteworthy that in recent years the AdS/CFT correspondence \cite{Maldacena:1998,Balasubramanian:2000} has provided many useful insights into the connection between different gauge theories and the gravity sector with the presence of various nonrotating black holes in AdS space. It would therefore be quite interesting to investigate the motion of test particles particularly around Reissner-Nordström-(anti-)de Sitter black holes surrounded by different kinds of regular and exotic matter fields to mark the impact of different fields like the quintessence on the structure of the orbits around them in diverse contexts.\\
In this paper we discuss the geodesic motion of test particles and light rays in the Reissner-Nordstr\"{o}m-(anti-)de Sitter spacetime in the presence of different regular and exotic matter fields. We provide our results here in terms of the Weierstra{\ss} $\wp$-function and in terms of derivatives of the Kleinian $\sigma$-function. The outline of this paper is as follows: In Sec. II we briefly review the metric of a Reissner-Nordstr\"{o}m-(anti-)de Sitter spacetime surrounded by various matter fields. In Sec. II we study possible orbits using parametric diagrams and effective potentials for four surrounding matter fields. In Sec. IV we present the analytical solution and plot some of the possible orbits in Sec. V. Sec. VI is dedicated to the conclusion.

\section{The spacetime structure}
In 2003 Kiselev \cite{Kiselev:2003} solved the Einstein-Maxwell field equation
\begin{equation}
G_{\mu\nu}+\Lambda g_{\mu\nu}=2 T_{\mu\nu}
\end{equation}  in the presence of a cosmological constant $\Lambda$. Here we use units where $c=4\pi G=1$. The additivity and linearity conditions on the quintessence dark energy stress tensor imply
\begin{equation}
	T_{t}^{t}=T_{r}^{r}=\rho_q
\end{equation}
\begin{equation}
	T_{\theta}^{\theta}=T_{\varphi}^{\varphi}=-\frac{\rho_q}{2}\left(3\omega_q+1\right)
\end{equation}
where, 
\begin{equation}
\rho_q=-\frac{a}{2}\frac{3\omega_q}{r^{3(\omega_q+1)}}\label{21}
\end{equation} 
and $\rho_q$ is the density of quintessence field, $\omega_q$ is the state parameter, and $a$ is the normalization parameter related to the density of quintessence field. In order to guarantee a positive density $\rho_q$ it was already pointed out in \cite{Kiselev:2003} that $a\omega_q\leq 0$. Also the unit of the normalization parameter $a$ is always the same as $r^{3\left(\omega_q+1\right)}$. Therefore the unit of $a$ changes for different types of surrounding matter. The metric of charged (anti-)de Sitter black hole immersed in quintessence dark energy as explained by Kiselev can be expressed by \cite{Kiselev:2003,Li:2014}
\begin{equation*}
ds^2=-Y(r)dt^2+\frac{dr^2}{Y(r)}+r^2\left(d\theta^2+\sin^2\theta\: d\varphi^2\right)\,,
\end{equation*}
with
\begin{equation}
Y(r)=1-\frac{2M}{r}+\frac{Q^2}{r^2}-\frac{a}{r^{3\omega_q+1}}-\frac{\Lambda}{3}r^2\,. \label{22} 
\end{equation}
Here $M=\frac{m}{4\pi}$ and $Q=\frac{q}{\sqrt{4\pi}}$, where $m$ is the usual Schwarzschild mass and $q$ the electric charge of the black hole. If we set $\Lambda=0$, $Q=0$ and put $\omega_q=-\frac{2}{3}$ the spacetime reduces to the Schwarzschild solution surrounded by quintessence whose geodesics were studied in \cite{Uniyal:2015}. Null trajectories of charged black holes surrounded by quintessence were carried out in \cite{Fernando:2015}; thermodynamic properties and the Joule-Thomson effect were investigated in \cite{Pradhan:2017} and \cite{Ghaffarnejad:2018}, respectively. The shadow of rotating charged black holes with quintessence was considered in \cite{Fernando:2015}. Other appropriate choices of $\omega_q$ allow us to obtain some well-known cases such as radiation field for $\omega_q=\frac{1}{3}$ and extraordinary quintessence for $\omega_q=-1$ named as the effective cosmological constant. In this Universe it could be possible that black holes are surrounded by regular $\omega_q=\frac{1}{3}$ as well as exotic matter fields, e.g., $\omega_q=\left[-\frac{1}{3}, -\frac{2}{3}, -1\right]$. Throughout this paper we restrict our analysis to these four cases, whereas two of them can be formulated as well-known special cases. Furthermore we only deal with spacetimes, where all of the black hole parameters are nonzero.
A crucial point of the spacetime structure is the existence of various numbers of horizons. For $\omega_q=-1$, we can define an effective cosmological constant $\Lambda'=\Lambda+a$. If this parameter is negative, then the spacetime reduces to the (anti-)de Sitter spacetime. Here the possible horizons are the event horizon, the Cauchy horizon and the cosmological horizon. If the set $\omega_q=-\frac{2}{3}$ an additional horizon can appear for a negative cosmological constant generated by the quintessence field. For $\omega=\pm\frac{1}{3}$ also up to three horizons are possible, whereas the $+$ sign corresponds to a Reissner-Nordstr\"{o}m-(anti-)de Sitter black hole with the effective charge $Q^2-a$.
\section{The Geodesic Equation}
In this section we derive the geodesic equation for the Reissner-Nordstr\"{o}m-(anti-)de Sitter black hole in the presence of various matter fields both for particles and light. To begin with we follow the standard Lagrangian procedure. The general form of the geodesic equation
\begin{equation}
\frac{d^2x^{\mu}}{d\lambda^{2}}+\Gamma^{\mu}_{\rho\sigma}\frac{d x^\rho}{d\lambda}\frac{d x^{\sigma}}{d\lambda}=0 \,,
\end{equation}
where $\Gamma^{\mu}_{\rho\sigma}$ is the Christoffel symbol of second kind and $\lambda$ is an affine parameter along the geodesic, which corresponds to the proper time for massive particles. The spacetime considered in this paper is given by the metric function in Eq. (\ref{22}). The Lagrangian reads
\begin{equation}
	\begin{aligned}
		\mathcal{L}&=\frac{1}{2}\left[-Y(r)\left(\frac{dt}{d\lambda}\right)^2+\frac{1}{Y(r)}\left(\frac{dr}{d\lambda}\right)^2+r^2\left(\frac{d\varphi}{d\lambda}\right)^2\right] \,,
	\end{aligned}
\end{equation}
where $\epsilon$ takes the value $1$ and $0$ for the massive and massless particle, respectively. Since the spacetime is spherically symmetric, we can restrict the analysis to the equatorial plane only. Furthermore using the Euler-Lagrange equation we obtain the constants of motion.
\begin{equation}
	\begin{aligned}
	 P_{t}&=\frac{\partial \mathcal{L}}{\partial\dot{t}}=-Y(r)\:\dot{t}=-E \\ \\
	 P_{\varphi}&=\frac{\partial \mathcal{L}}{\partial\dot{\varphi}}=r^2\dot{\varphi}=L \,,
	\end{aligned}
\end{equation}
where $E$ is the energy and $L$ is the angular momentum of the particle. Using the above conserved quantities the geodesic equation can be written as an ordinary differential equation that involves $r$ as a function of $\varphi$
\begin{equation}
	\left(\frac{dr}{d\varphi}\right)^2=\frac{r^4}{L^2}\left[E^2-\left(\epsilon+\frac{L^2}{r^2}\right)\left(1-\frac{2M}{r}+\frac{Q^2}{r^2}-\frac{a}{r^{3\omega_q+1}}-\frac{\Lambda}{3}r^2\right)\right]\,. \label{23} 
\end{equation}
To obtain the turning points of the geodesic motion, an effective potential can be introduced easily
\begin{equation}
	V=\left(\epsilon+\frac{L^2}{r^2}\right)\left(1-\frac{2M}{r}+\frac{Q^2}{r^2}-\frac{a}{r^{3\omega_q+1}}-\frac{\Lambda}{3}r^2\right) \,.
\label{eqn:Veff}
\end{equation}
For a simplified analysis of the geodesic equation it is convenient to introduce dimensionless quantities as follows
\begin{equation*}
	\tilde{r}=\frac{r}{M},\quad\tilde{L}=\frac{M^2}{L^2},\quad\tilde{Q}=\frac{Q}{M},\quad\tilde{a}=\frac{a}{M^{3\omega_q+1}},\quad\tilde{\Lambda}=\frac{\Lambda M^2}{3}\,.
\end{equation*}
Now the scaled form of Eq. \ref{23} is given by
\begin{multline}
	\left(\frac{d\tilde{r}}{d\varphi}\right)^2=\epsilon \tilde{\Lambda}\tilde{L}\tilde{r}^6+\left((E^2-\epsilon)\tilde{L}+\tilde{\Lambda}\right)\tilde{r}^4+2\epsilon\tilde{L}\tilde{r}^3\\-\left(\epsilon\tilde{L}\tilde{Q}^2+1\right)\tilde{r}^2+2\tilde{r}-\tilde{Q}^2+\frac{\epsilon\tilde{a}\tilde{L}}{\tr^{3\omega_q-3}}+\frac{\tilde{a}}{\tr^{3\omega_q-1}}=R(\tilde{r})\,. \label{6}
\end{multline}
The above equation implies that $R(\tilde{r})\ge0$ is a necessary condition for the existence of geodesic motion and thus the real $0$'s of $R(\tilde{r})$ are the turning points of the geodesic motion. In the forthcoming chapters we will study the analytic solutions of the geodesic equation obtained by varying the state parameter $\omega_q$.

\subsection{Preliminary discussion on the possible orbit types}
Equation. (\ref{6}) suggests that the shape of an orbit depends on the energy $E$ and the angular momentum $\tL$ of the test particle or light ray as well as the state parameter $\omega_q$ and the normalization parameter $\tilde{a}$. The physically acceptable regions of the geodesic motion are given by the real values of $\tr$ for which $R(\tilde{r})\ge0$ or equivalently $E^2>V_{eff}$. In the following we present a list of the possible orbit types. We consider $\tilde{r}_{+}$ and $\tilde{r}_{-}$ as the Event and Cauchy horizon respectively such that $\tr_+>\tr_-$. We distinguish between the following orbit types
\begin{enumerate}
	\item Escape orbit (EO) with range $\tilde{r}\in[\tilde{r}_1,\infty)$, where $\tilde{r}_1>\tilde{r}_{+}$.
	
	\item Bound orbit (BO) with range $\tilde{r}\in[\tilde{r}_1,\tilde{r}_2]$, where $\tilde{r}_1,\tilde{r}_2>\tilde{r}_{+}$.
	
	\item Two-world escape orbit (TEO) with range $\tilde{r}\in[\tilde{r}_1,\infty)$, where $0<\tilde{r}_1<\tilde{r}_{-}$.
	
	\item Many-world bound orbit (MBO) with range $\tilde{r}\in[\tilde{r}_1,\tilde{r}_2]$, where $0<\tilde{r}_1\le \tilde{r}_{-}$ and $\tilde{r}_2\ge \tilde{r}_{+}$.
\end{enumerate}
 
\subsection{Parametric diagrams}
\begin{figure}[H]
	\centering
	\subfigure[$\epsilon=1$, $\ta=0.8$, $\tLambda=-0.001$ and $\tQ=0.1$]{
		 \includegraphics[width=0.45\textwidth]{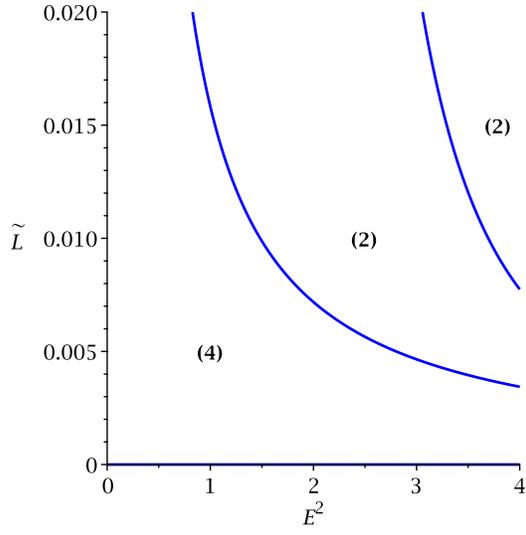}
	}
	\subfigure[$\epsilon=0$, $\ta=0.8$, $\tLambda=-0.001$ and $\tQ=0.1$]{
		\includegraphics[width=0.45\textwidth]{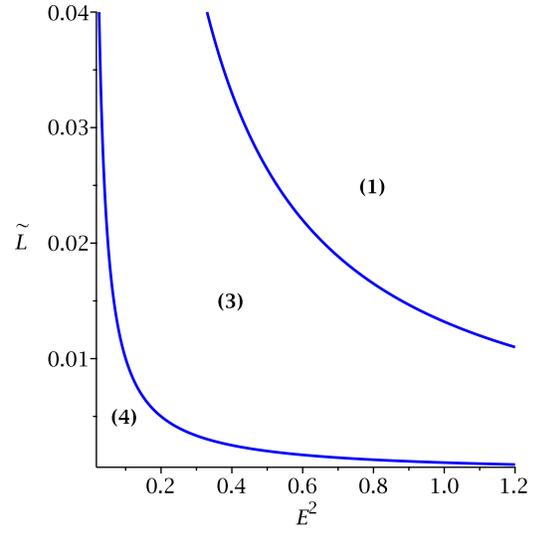}
	}
	
	\subfigure[$\epsilon=1$, $\ta=0.1$, $\tLambda=0.001$ and $\tQ=0.1$]{
		\includegraphics[width=0.45\textwidth]{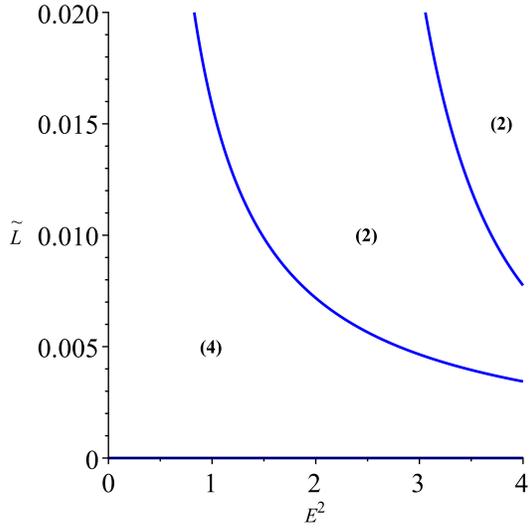}
	}
	\subfigure[$\epsilon=0$, $\ta=0.1$, $\tLambda=0.001$ and $\tQ=0.1$]{
		\includegraphics[width=0.45\textwidth]{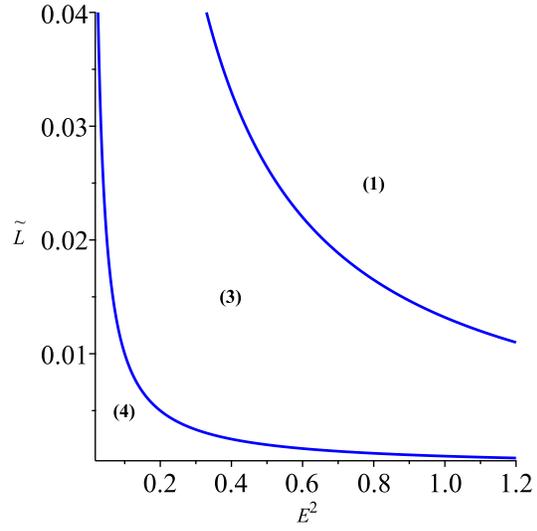}
	}
	\caption{Parametric $\tL$-$E^2$ diagrams for particles and light for the state parameter $\omega_q=-\frac{2}{3}$. The numbers in brackets denote the number of positive real $0$'s of the polynomial $R$.}
	\label{pic:param_066}
\end{figure}

\begin{figure}[H]
	\centering
	\subfigure[$\epsilon=1$, $\ta=0.8$, $\tLambda=-0.001$ and $\tQ=0.15$]{
		 \includegraphics[width=0.45\textwidth]{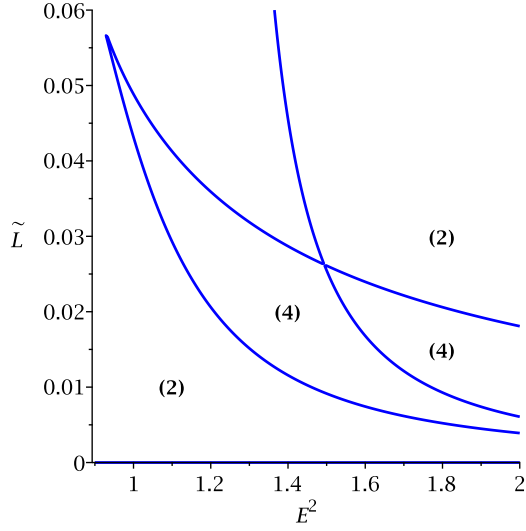}
	}
	\subfigure[$\epsilon=0$, $\ta=0.8$, $\tLambda=-0.001$ and $\tQ=0.15$]{
		\includegraphics[width=0.45\textwidth]{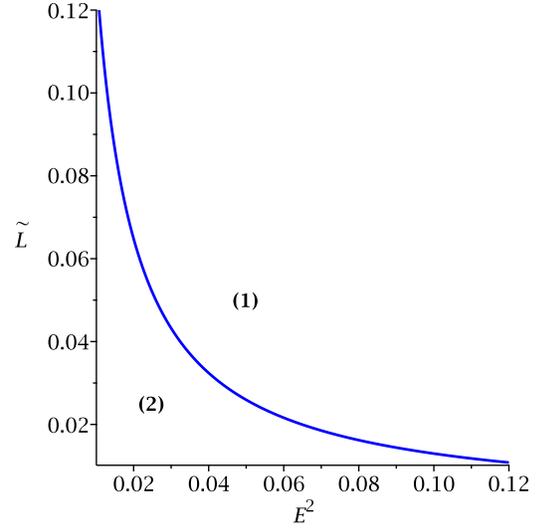}
	}
	
\subfigure[$\epsilon=1$, $\ta=0.1$, $\tLambda=0.001$ and $\tQ=0.15$]{
		 \includegraphics[width=0.45\textwidth]{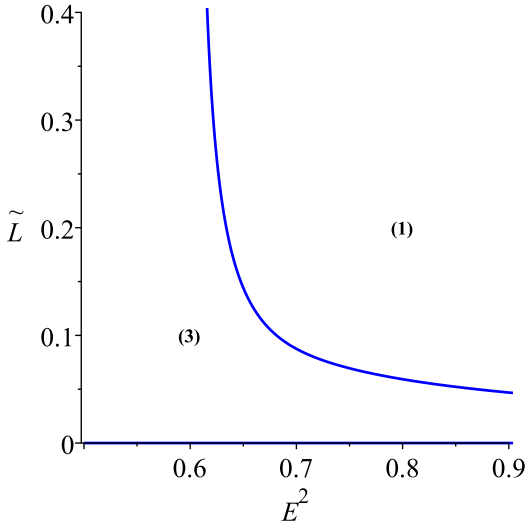}
	}
	\subfigure[$\epsilon=0$, $\ta=0.1$, $\tLambda=0.001$ and $\tQ=0.15$]{
		\includegraphics[width=0.45\textwidth]{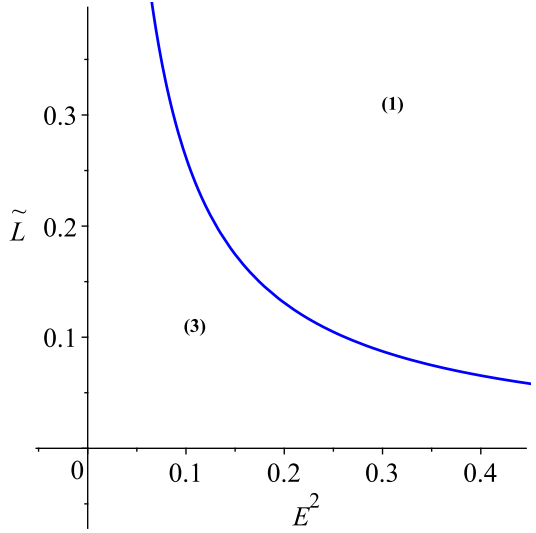}
	}
	\caption{Parametric $\tL$-$E^2$ diagrams for particles and light for the state parameter $\omega_q=-\frac{1}{3}$. The numbers in brackets denote the number of positive real $0$'s of the polynomial $R$.}
	\label{pic:param_033}
\end{figure}

The $\tilde{r}-\varphi$ equation for each kind of surrounding matter fields can be obtained from Eq. (\ref{6}) through a suitable choice of $\omega_q$. If for a given set of parameters $\tilde{\Lambda}$, $\tilde{a}$, $\tilde{Q}$, $\epsilon$, $E^2$, $\tilde{L}$, the polynomial (\ref{6}) has $n$ positive real $0$'s, then for varying $E^2$ and $\tilde{L}$ this number can only change if two $0$'s merge to one. Solving $R(\tilde{r})=0$ and $\frac{dR(\tilde{r})}{d\tilde{r}}=0$ for $E^2$ and $\tilde{L}$ for different matter fields we obtain the parametric diagrams shown in Figs. \ref{pic:param_066} and \ref{pic:param_033}.

\subsection{Effective potentials}
\begin{figure}[H]
	\centering
	\subfigure[$\epsilon=1$, $\ta=0.03$, $\tLambda=-0.01$, $\tQ=0.1$ and $\tL=0.04$]{
		 \includegraphics[width=0.45\textwidth]{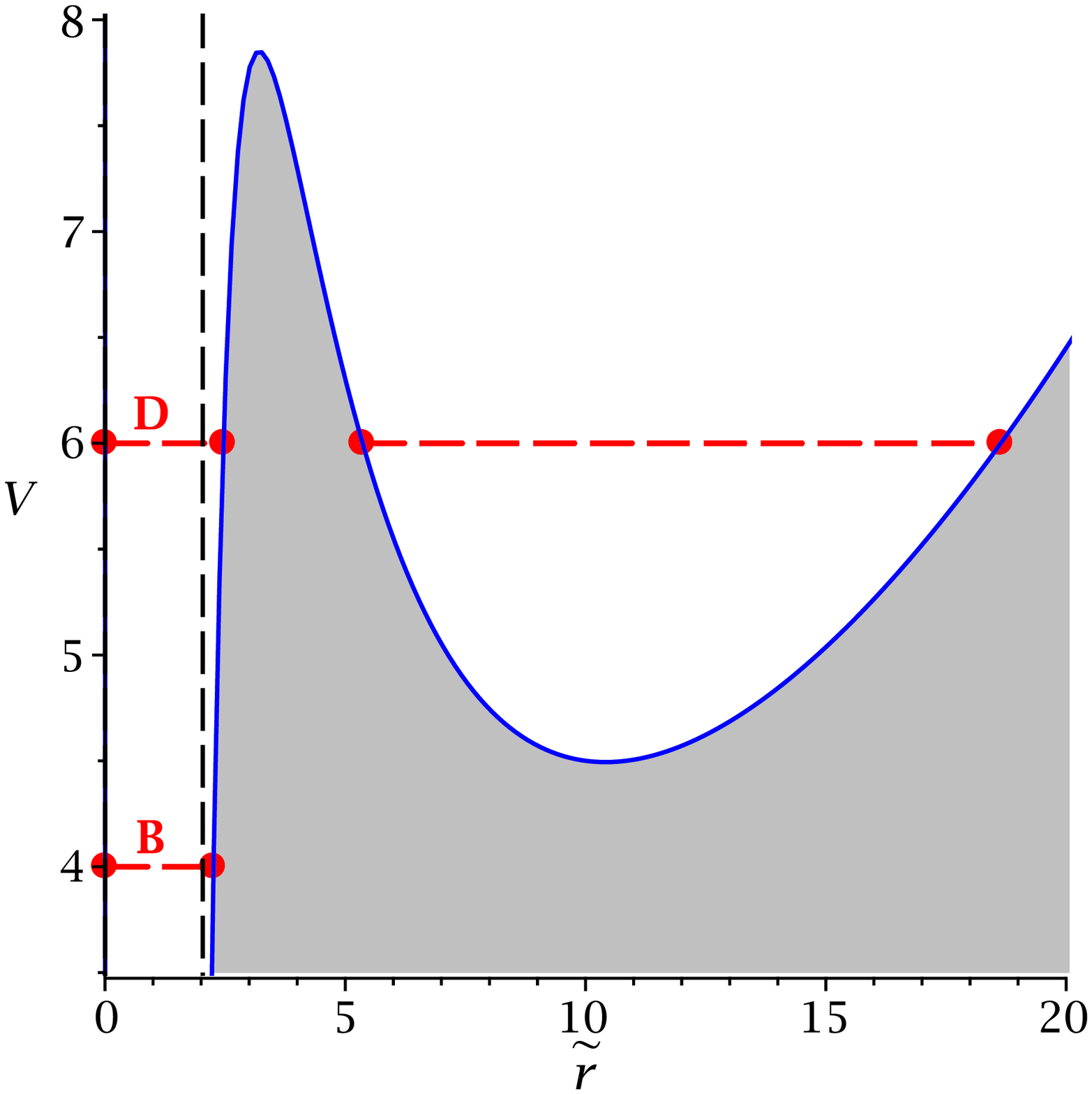}
	}
	\subfigure[$\epsilon=0$, $\ta=0.03$, $\tLambda=-0.01$, $\tQ=0.1$ and $\tL=0.04$]{
		\includegraphics[width=0.45\textwidth]{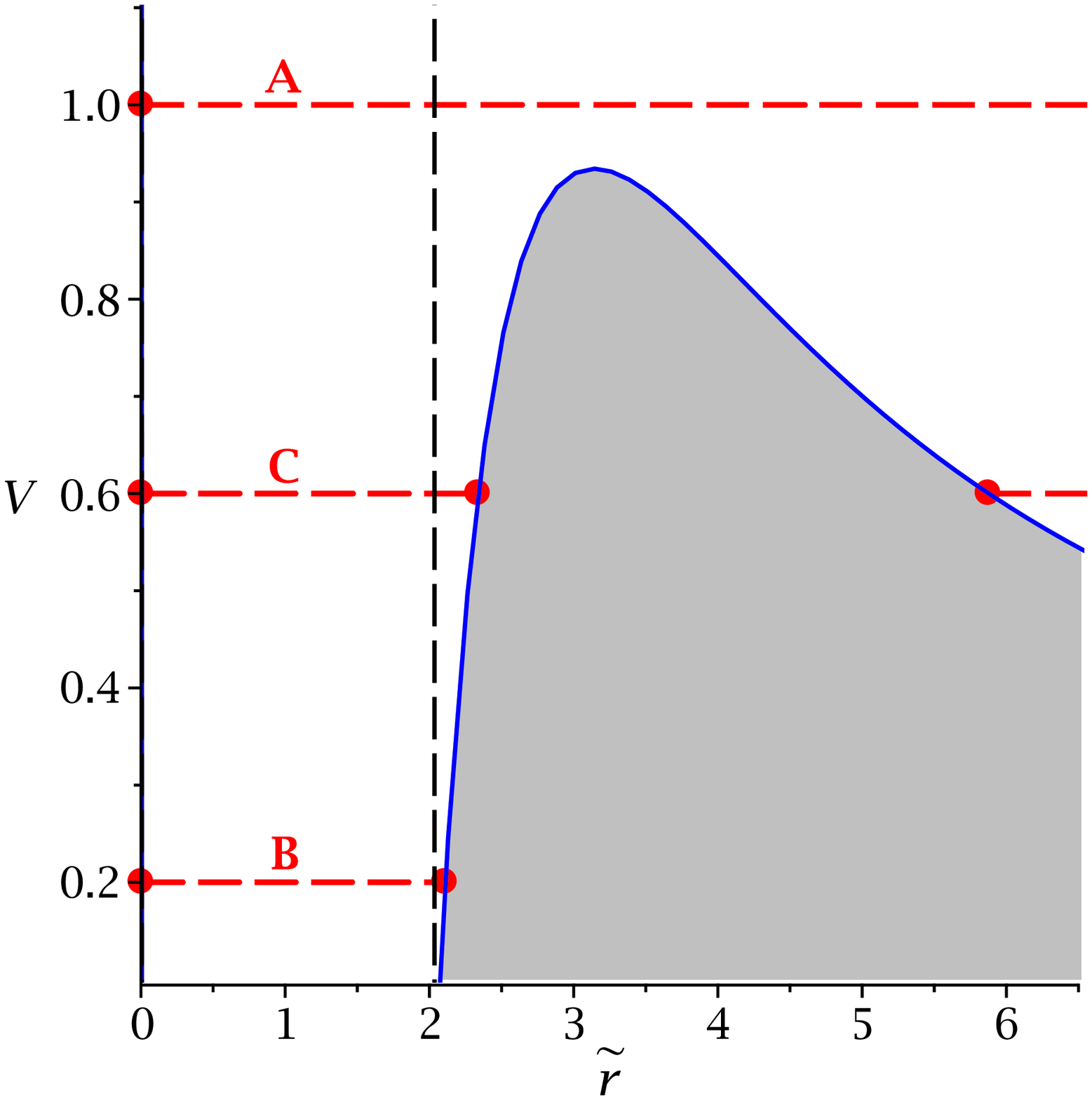}
	}
	
	\subfigure[$\epsilon=1$, $\ta=0.03$, $\tLambda=0.01$, $\tQ=0.1$ and $\tL=0.005$]{
		 \includegraphics[width=0.45\textwidth]{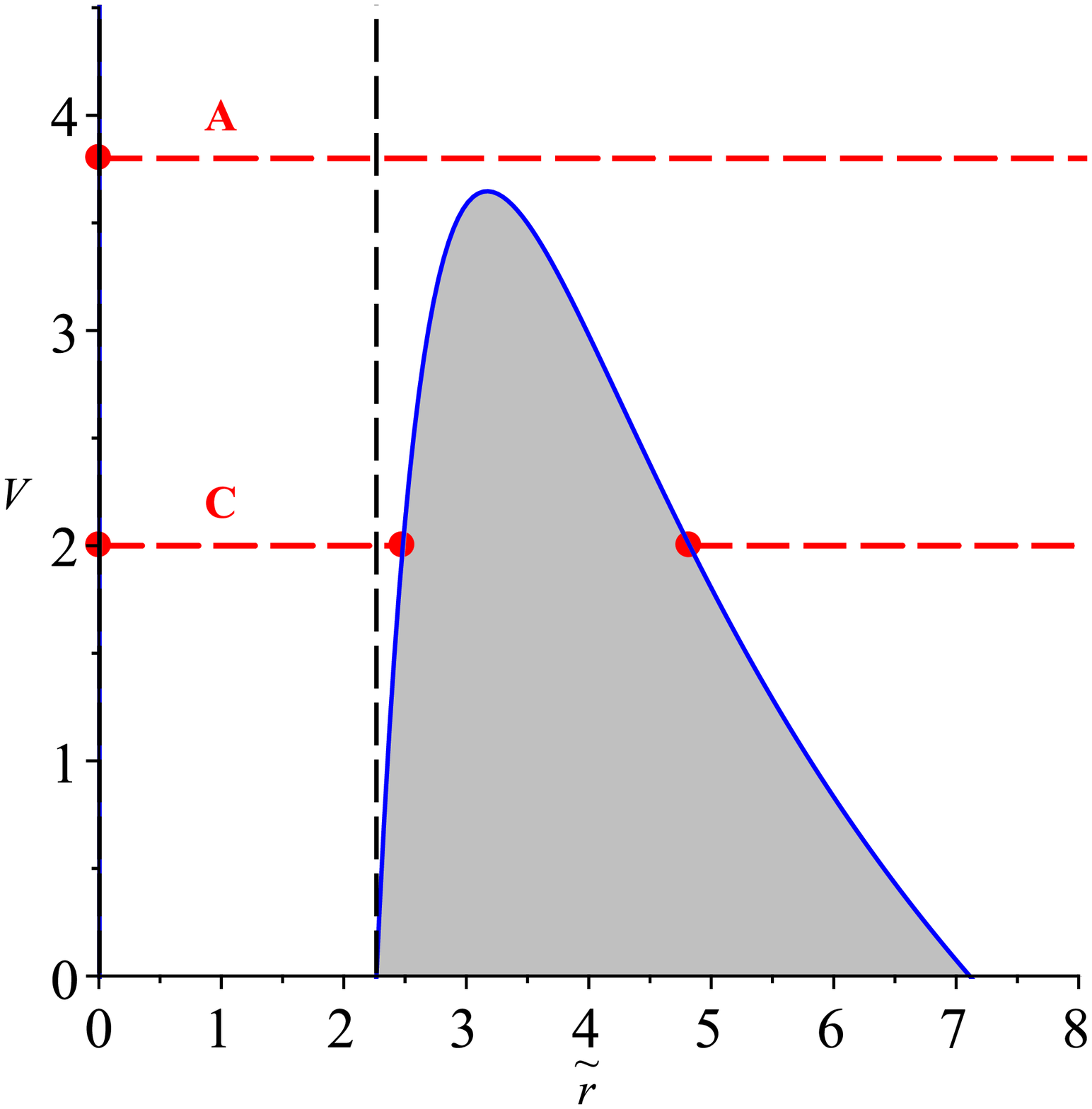}
	}
	\subfigure[$\epsilon=0$, $\ta=0.03$, $\tLambda=0.01$, $\tQ=0.1$ and $\tL=0.04$]{
		\includegraphics[width=0.45\textwidth]{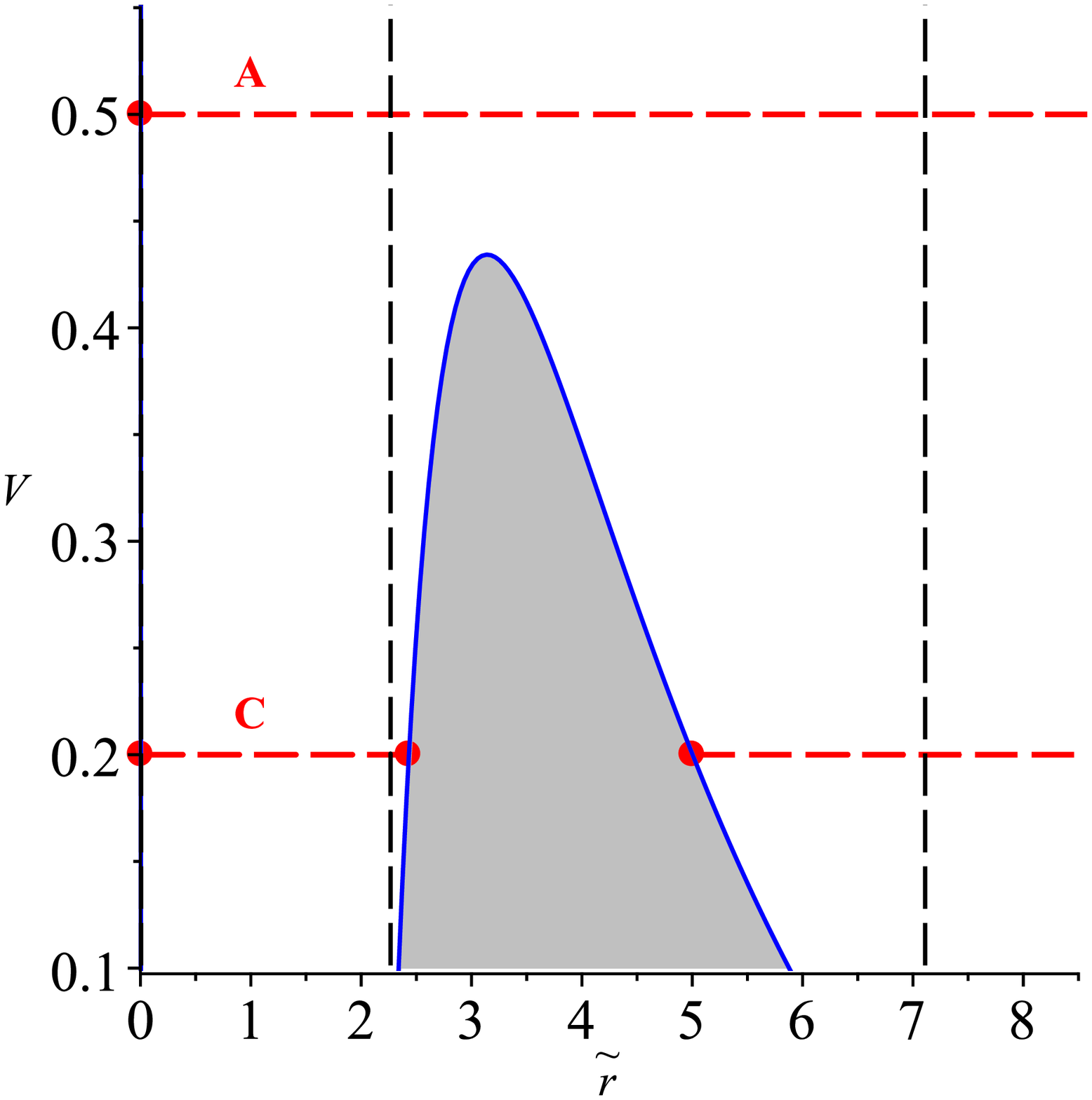}
	}
	
\subfigure[$\epsilon=1$, $\ta=0.1$, $\tLambda=-0.0025$, $\tQ=0.005$ and $\tL=0.2$]{
		 \includegraphics[width=0.45\textwidth]{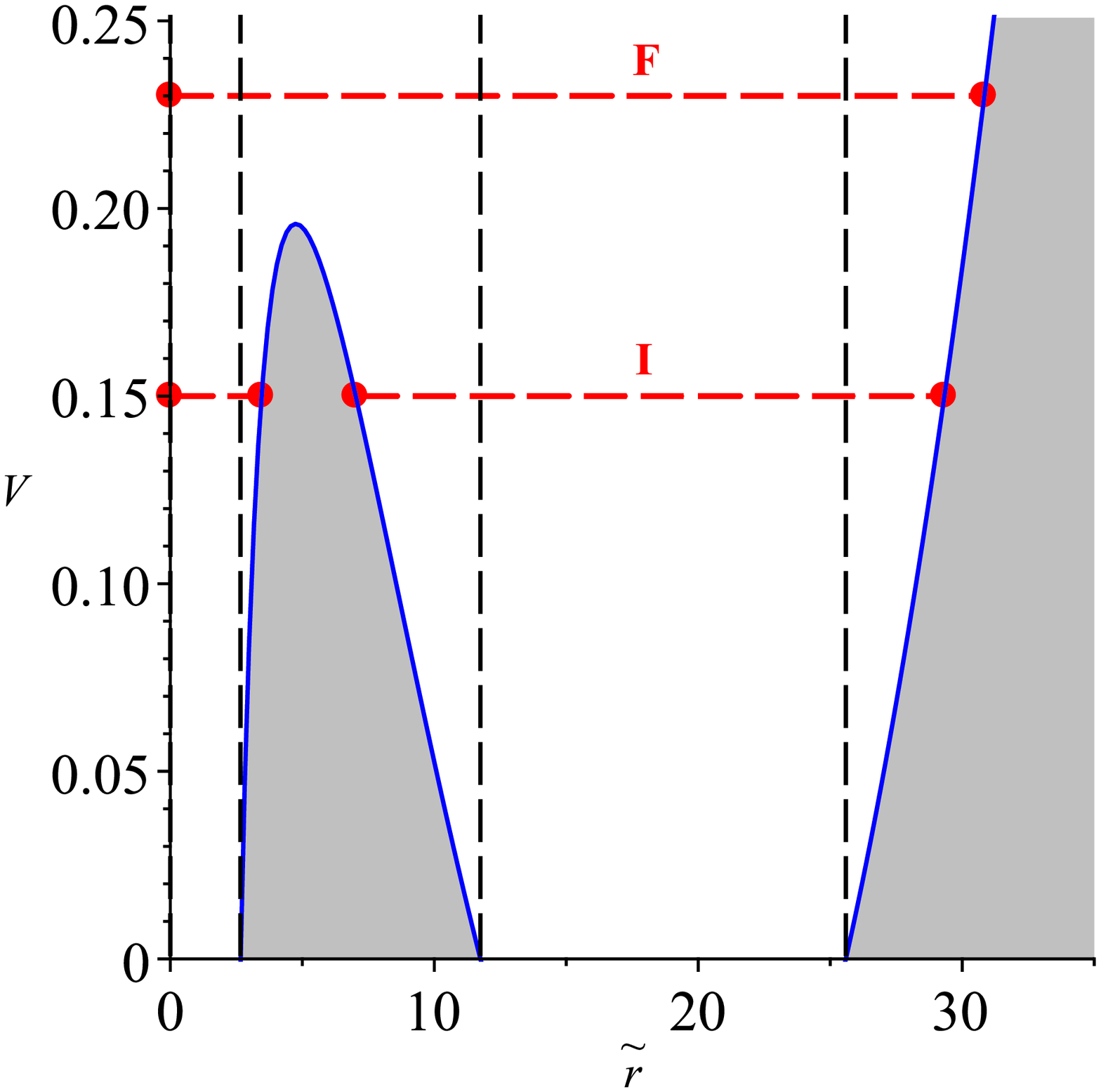}
	}
	\subfigure[$\epsilon=0$, $\ta=0.1$, $\tLambda=-0.0025$, $\tQ=0.005$ and $\tL=0.08$]{
		\includegraphics[width=0.45\textwidth]{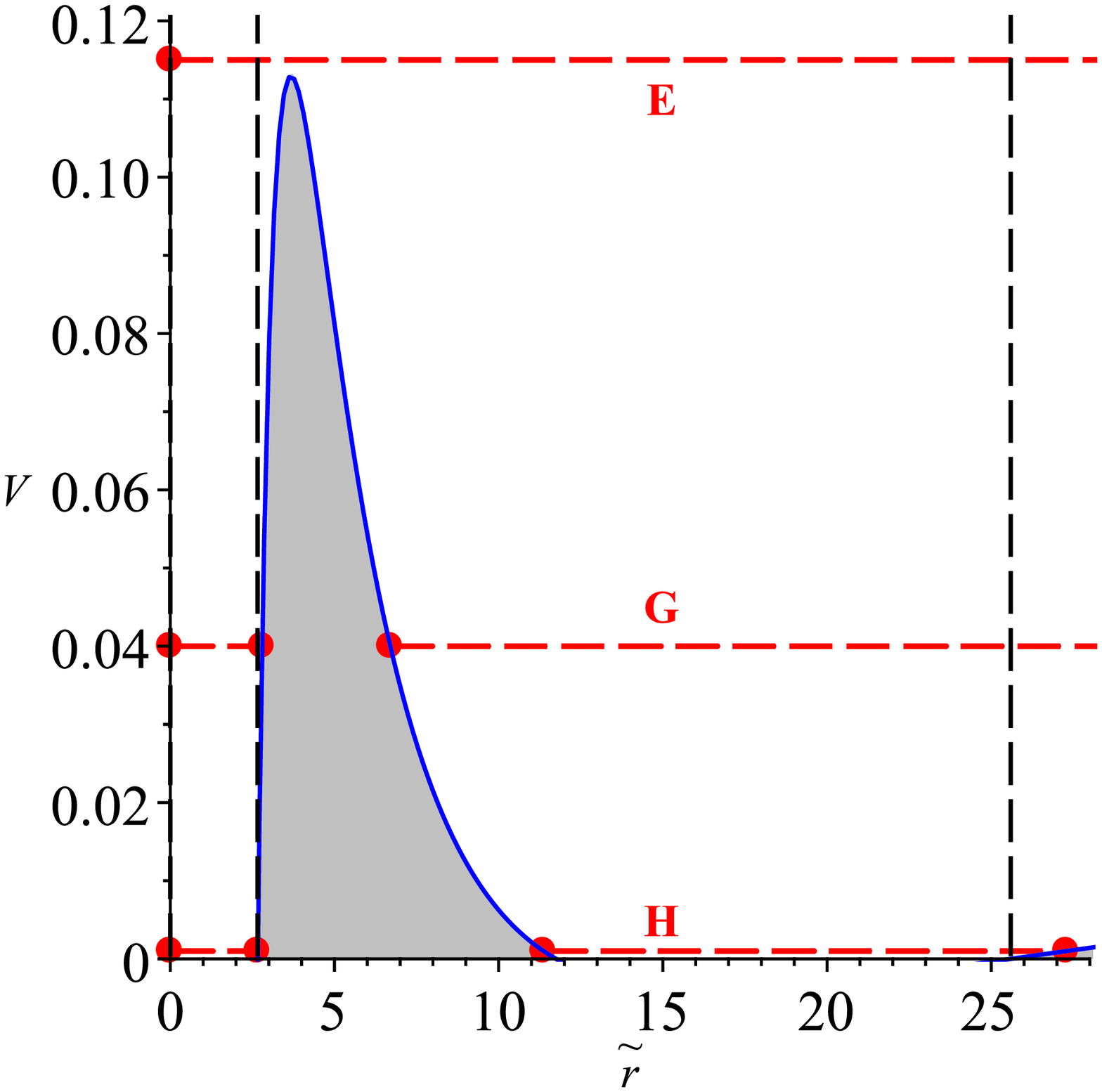}
	}
	\caption{Effective potential $V$ and various energies for different kinds of orbits for the state parameter $\omega_q=-\frac{2}{3}$. The blue line shows the potential and the grey area is forbidden, because here $R(\tr)$ becomes negative. The turning points of the particle are denoted by red dots.}
	\label{pic:potential_066}
\end{figure}

\begin{figure}[H]
	\centering
	\subfigure[$\epsilon=1$, $\ta=0.08$, $\tLambda=-0.001$, $\tQ=0.15$ and $\tL=0.03$]{
		 \includegraphics[width=0.45\textwidth]{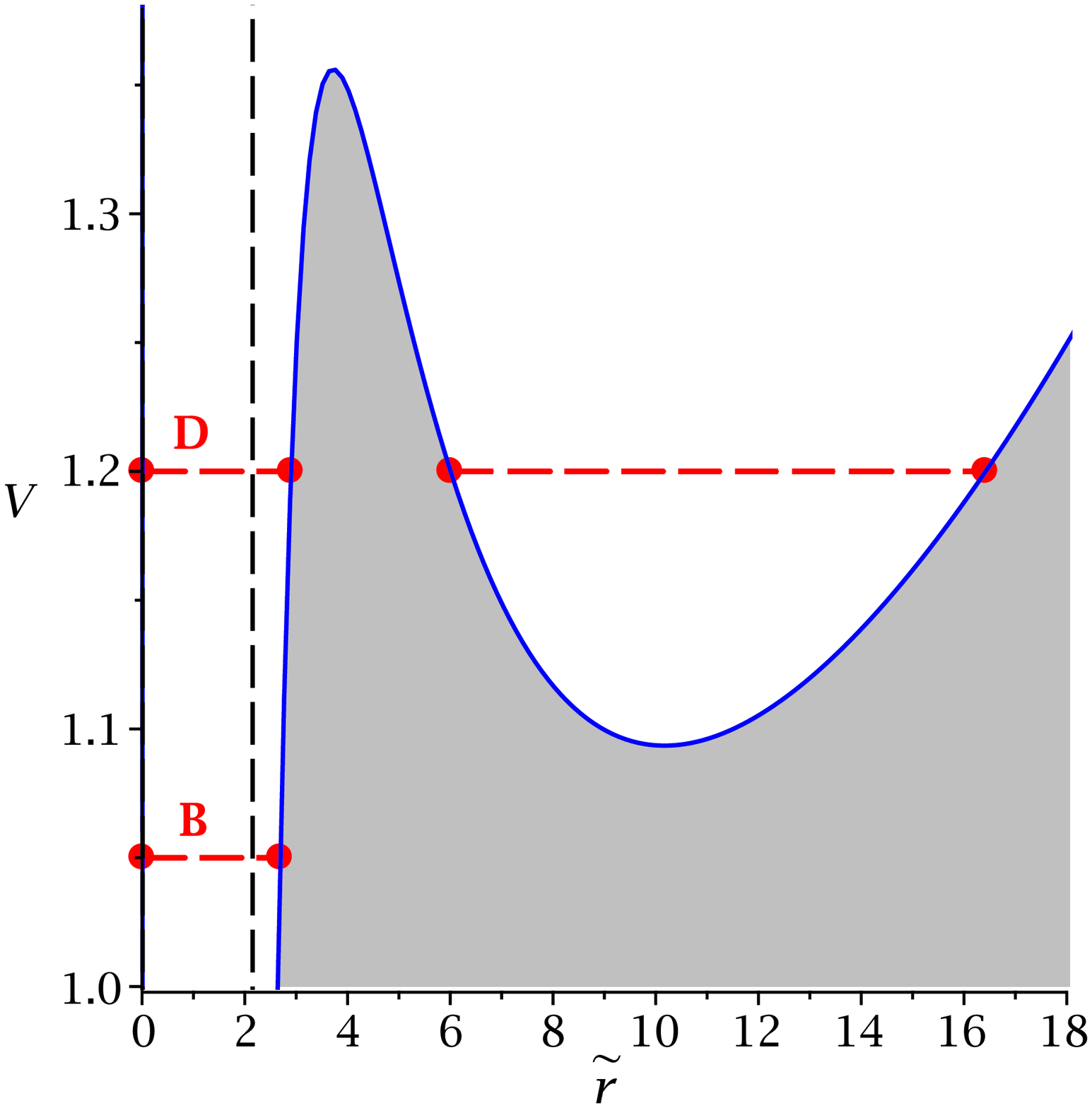}
	}
	\subfigure[$\epsilon=0$, $\ta=0.08$, $\tLambda=-0.01$, $\tQ=0.15$ and $\tL=0.1$]{
		\includegraphics[width=0.45\textwidth]{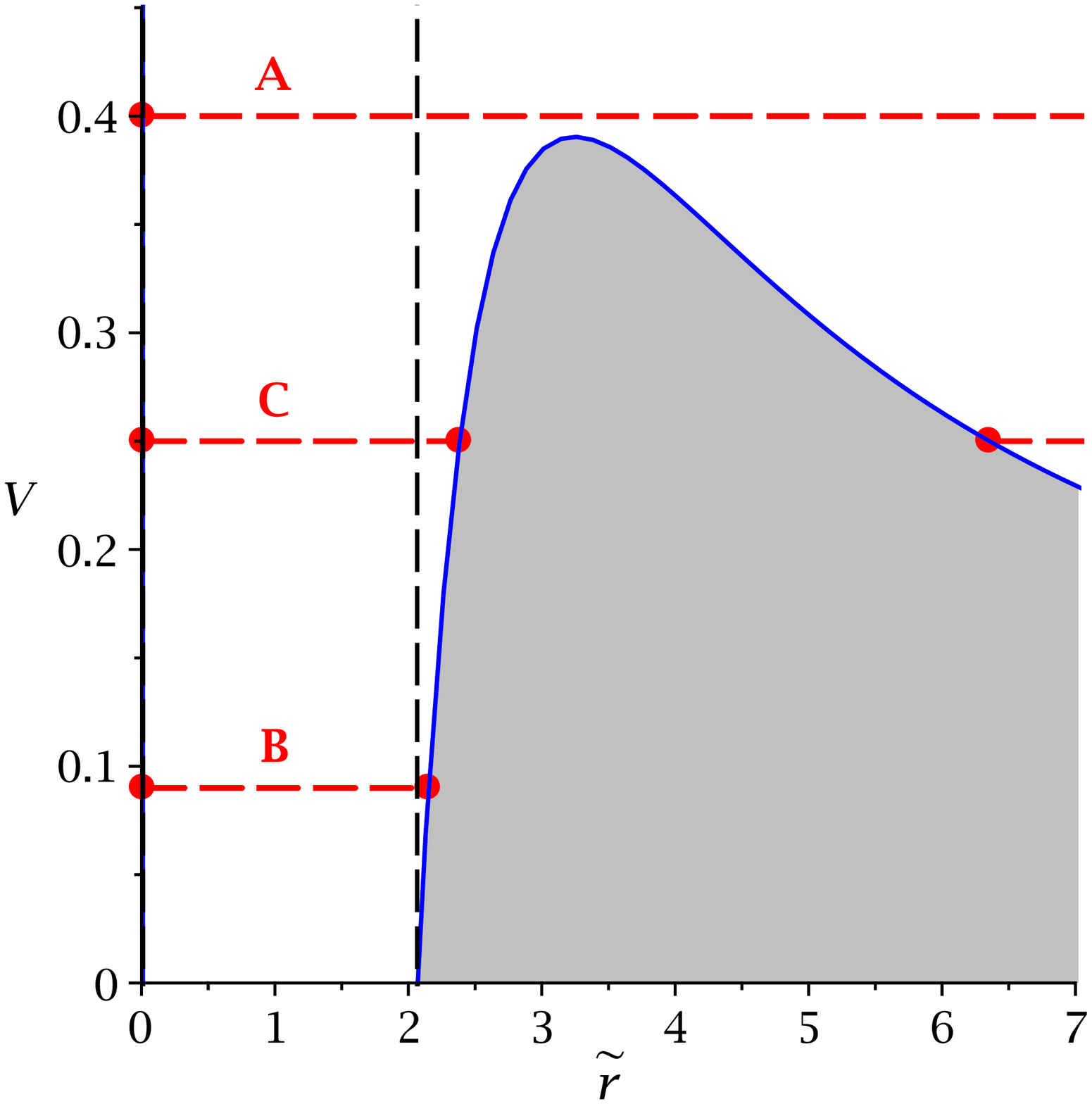}
	}
	
	\subfigure[$\epsilon=1$, $\ta=0.1$, $\tLambda=0.001$, $\tQ=0.15$ and $\tL=0.2$]{
		 \includegraphics[width=0.45\textwidth]{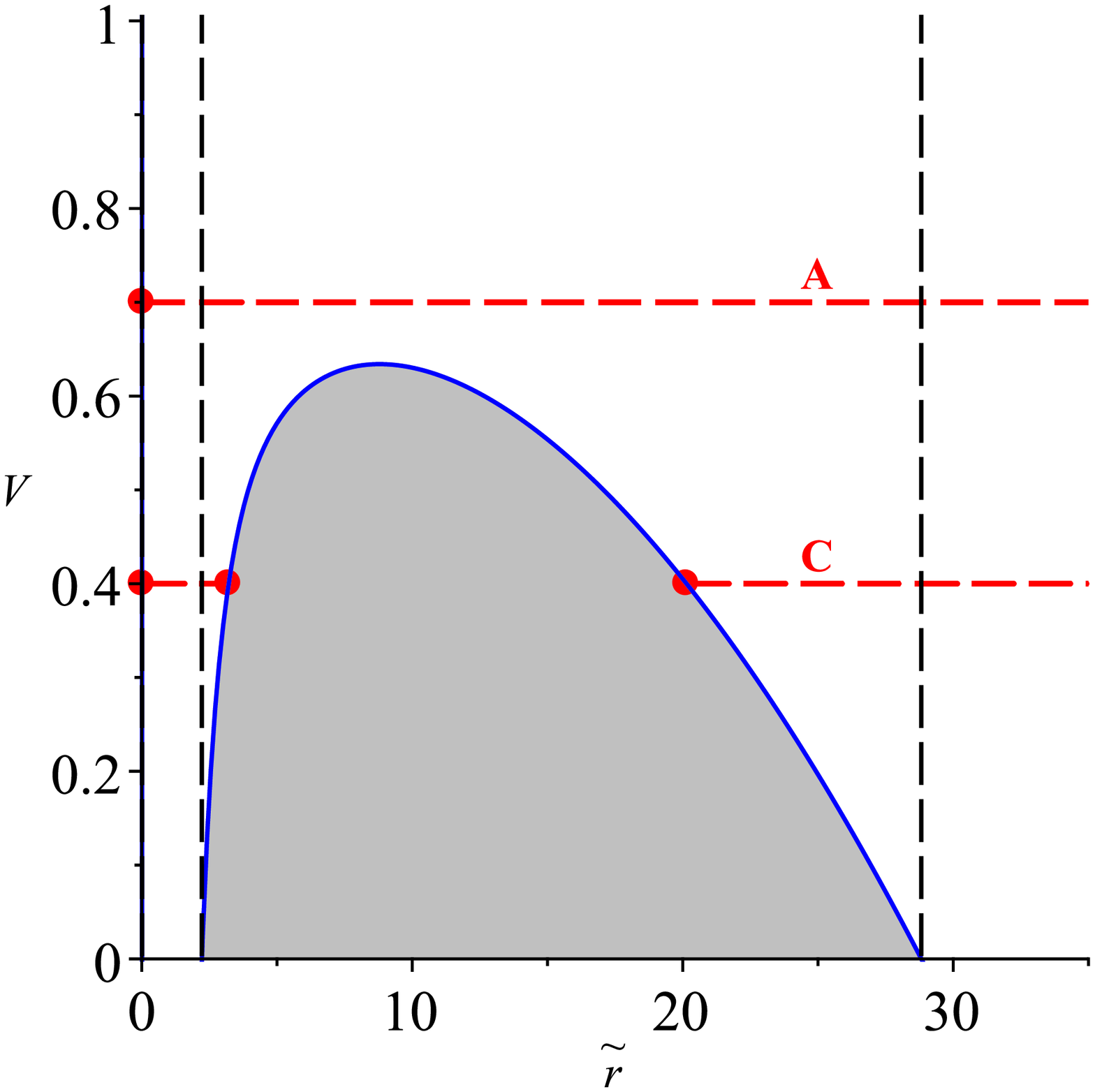}
	}
	\subfigure[$\epsilon=0$, $\ta=0.1$, $\tLambda=0.001$, $\tQ=0.15$ and $\tL=0.6$]{
		\includegraphics[width=0.45\textwidth]{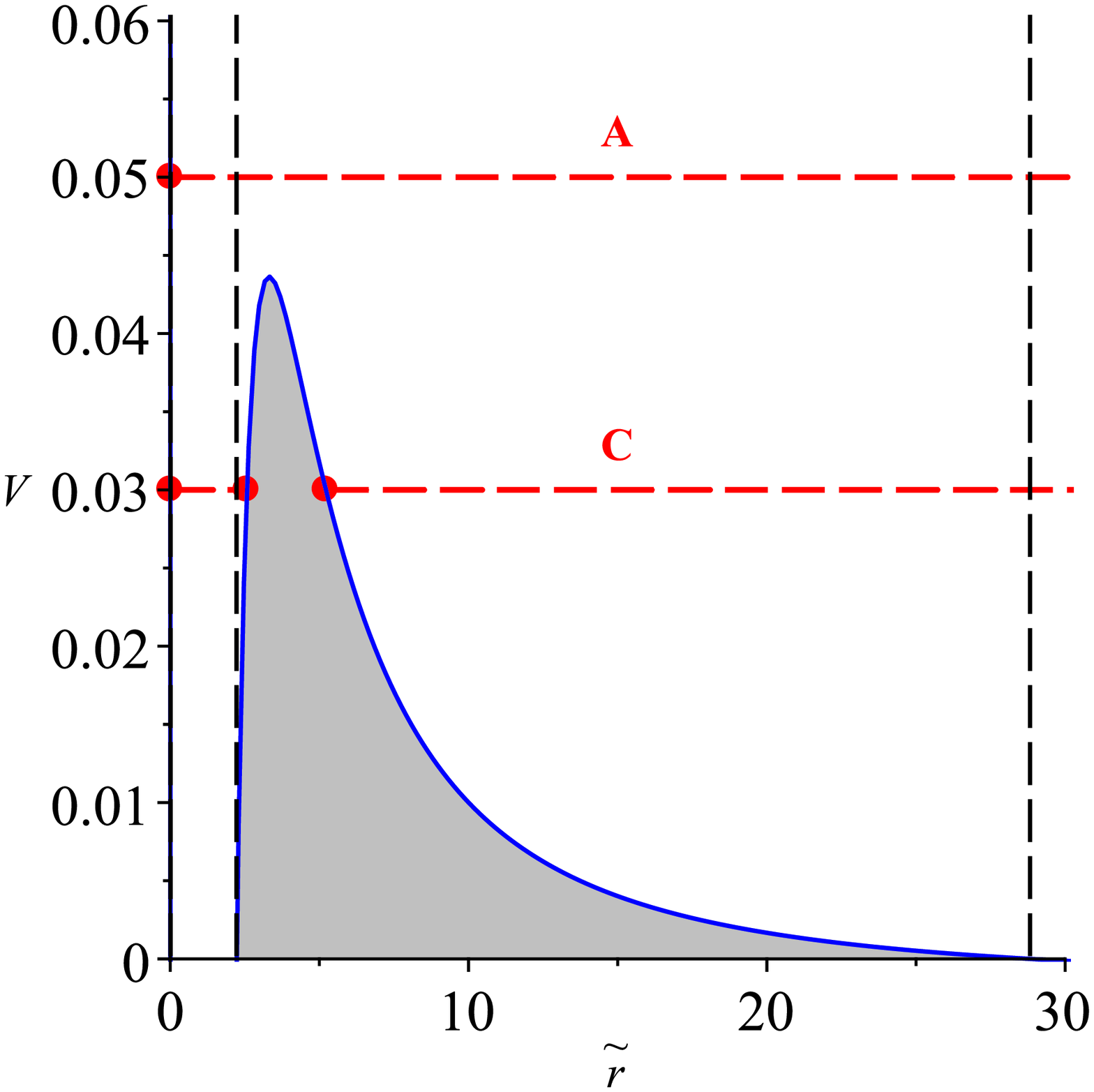}
	}
	\caption{Effective potential $V$ and various energies for different kinds of orbits for the state parameter $\omega_q=-\frac{1}{3}$. The blue line shows the potential and the grey area is forbidden, because here $R(\tr)$ becomes negative. The turning points of the particle are denoted by red dots.}
	\label{pic:potential_033}
\end{figure}

Another possibility to determine the location of the $0$'s of the polynomial $R(\tr)$ (e.g. the turning points of the geodesic motion) is to use the effective potential $V$ defined in Eq. \ref{eqn:Veff}. In Figs. \ref{pic:potential_066} and \ref{pic:potential_033} we illustrated the effective potential for the two values of $\omega_q$ each for test particles and light. Here the intersections of the different particle energies $E$ with the effective potential correspond to the turning points of the particle motion.
Combining the parametric diagrams and the effective potentials we can present a list of all possible orbits for each value of $\omega_q$ in Tables \ref{tab:orbit-types_066} and \ref{tab:orbit-types_033}. 
\begin{table}[H]
\begin{center}
\begin{tabular}{|lccll|}\hline
Type &  $0$'s & $\epsilon_{sign\left(\Lambda\right)}$   & Range of $\tr$ & Orbit \\
\hline\hline
  A & 1 & $0_{\pm}$   &
\begin{pspicture}(-4,-0.2)(3.5,0.2)
\psline[linewidth=0.5pt]{-}(-4,0)(3.5,0)
\pscircle[hatchcolor=white,fillstyle=solid](-2,0){0.075}
\psline[linewidth=0.5pt,doubleline=true](-0.9,-0.2)(-0.9,0.2)
\psline[linewidth=0.5pt,doubleline=true](-0.1,-0.2)(-0.1,0.2)
\psline[linewidth=0.5pt,doubleline=true](2.2,-0.2)(2.2,0.2)
\psline[linewidth=1.2pt]{*-}(-1.2,0)(3.5,0)
\end{pspicture}
& TEO 
\\ \hline
B & 2 & $1_{\pm}$, $0_{\pm}$  &
\begin{pspicture}(-4,-0.2)(3.5,0.2)
\psline[linewidth=0.5pt]{-}(-4,0)(3.5,0)
\pscircle[hatchcolor=white,fillstyle=solid](-2,0){0.075}
\psline[linewidth=0.5pt,doubleline=true](-0.9,-0.2)(-0.9,0.2)
\psline[linewidth=0.5pt,doubleline=true](-0.1,-0.2)(-0.1,0.2)
\psline[linewidth=0.5pt,doubleline=true](2.2,-0.2)(2.2,0.2)
\psline[linewidth=1.2pt]{*-*}(-1.2,0)(0.2,0)
\end{pspicture}
& MBO
\\ \hline
C &3 & $1_{+}$, $0_{+}$  &
\begin{pspicture}(-4,-0.2)(3.5,0.2)
\psline[linewidth=0.5pt]{-}(-4,0)(3.5,0)
\pscircle[hatchcolor=white,fillstyle=solid](-2,0){0.075}
\psline[linewidth=0.5pt,doubleline=true](-0.9,-0.2)(-0.9,0.2)
\psline[linewidth=0.5pt,doubleline=true](-0.1,-0.2)(-0.1,0.2)
\psline[linewidth=0.5pt,doubleline=true](2.2,-0.2)(2.2,0.2)
\psline[linewidth=1.2pt]{*-*}(-1.2,0)(0.2,0)
\psline[linewidth=1.2pt]{*-}(1,0)(3.5,0)
\end{pspicture}
  &  MBO, TEO
\\ \hline
D & 4 & $1_{\pm}$, $0_{\pm}$  &
\begin{pspicture}(-4,-0.2)(3.5,0.2)
\psline[linewidth=0.5pt]{-}(-4,0)(3.5,0)
\pscircle[hatchcolor=white,fillstyle=solid](-2,0){0.075}
\psline[linewidth=0.5pt,doubleline=true](-0.9,-0.2)(-0.9,0.2)
\psline[linewidth=0.5pt,doubleline=true](-0.1,-0.2)(-0.1,0.2)
\psline[linewidth=0.5pt,doubleline=true](2.2,-0.2)(2.2,0.2)
\psline[linewidth=1.2pt]{*-*}(-1.2,0)(0.2,0)
\psline[linewidth=1.2pt]{*-*}(1,0)(1.8,0)
\end{pspicture}
  &  MBO, BO
\\ \hline
E & 1 &  $0_{-}$  &
\begin{pspicture}(-4,-0.2)(3.5,0.2)
\psline[linewidth=0.5pt]{-}(-4,0)(3.5,0)
\pscircle[hatchcolor=white,fillstyle=solid](-2,0){0.075}
\psline[linewidth=0.5pt,doubleline=true](-0.9,-0.2)(-0.9,0.2)
\psline[linewidth=0.5pt,doubleline=true](-0.1,-0.2)(-0.1,0.2)
\psline[linewidth=0.5pt,doubleline=true](1.4,-0.2)(1.4,0.2)
\psline[linewidth=0.5pt,doubleline=true](2.2,-0.2)(2.2,0.2)
\psline[linewidth=1.2pt]{*-}(-1.2,0)(3.5,0)
\end{pspicture}
  &  TEO
\\ \hline
F & 2 &  $1_{-}$  &
\begin{pspicture}(-4,-0.2)(3.5,0.2)
\psline[linewidth=0.5pt]{-}(-4,0)(3.5,0)
\pscircle[hatchcolor=white,fillstyle=solid](-2,0){0.075}
\psline[linewidth=0.5pt,doubleline=true](-0.9,-0.2)(-0.9,0.2)
\psline[linewidth=0.5pt,doubleline=true](-0.1,-0.2)(-0.1,0.2)
\psline[linewidth=0.5pt,doubleline=true](1.4,-0.2)(1.4,0.2)
\psline[linewidth=0.5pt,doubleline=true](2.2,-0.2)(2.2,0.2)
\psline[linewidth=1.2pt]{*-*}(-1.2,0)(2.5,0)
\end{pspicture}
  &  MBO
\\ \hline
G & 3 &  $0_{-}$  &
\begin{pspicture}(-4,-0.2)(3.5,0.2)
\psline[linewidth=0.5pt]{-}(-4,0)(3.5,0)
\pscircle[hatchcolor=white,fillstyle=solid](-2,0){0.075}
\psline[linewidth=0.5pt,doubleline=true](-0.9,-0.2)(-0.9,0.2)
\psline[linewidth=0.5pt,doubleline=true](-0.1,-0.2)(-0.1,0.2)
\psline[linewidth=0.5pt,doubleline=true](1.4,-0.2)(1.4,0.2)
\psline[linewidth=0.5pt,doubleline=true](2.2,-0.2)(2.2,0.2)
\psline[linewidth=1.2pt]{*-*}(-1.2,0)(0.2,0)
\psline[linewidth=1.2pt]{*-}(1,0)(3.5,0)
\end{pspicture}
  &  MBO, EO
\\ \hline
H & 4 &  $0_{-}$  &
\begin{pspicture}(-4,-0.2)(3.5,0.2)
\psline[linewidth=0.5pt]{-}(-4,0)(3.5,0)
\pscircle[hatchcolor=white,fillstyle=solid](-2,0){0.075}
\psline[linewidth=0.5pt,doubleline=true](-0.9,-0.2)(-0.9,0.2)
\psline[linewidth=0.5pt,doubleline=true](-0.1,-0.2)(-0.1,0.2)
\psline[linewidth=0.5pt,doubleline=true](1.4,-0.2)(1.4,0.2)
\psline[linewidth=0.5pt,doubleline=true](2.2,-0.2)(2.2,0.2)
\psline[linewidth=1.2pt]{*-*}(-1.2,0)(0.2,0)
\psline[linewidth=1.2pt]{*-*}(0.45,0)(1,0)
\end{pspicture}
  &  MBO, BO
\\ \hline
I & 4 &  $1_{-}$  &
\begin{pspicture}(-4,-0.2)(3.5,0.2)
\psline[linewidth=0.5pt]{-}(-4,0)(3.5,0)
\pscircle[hatchcolor=white,fillstyle=solid](-2,0){0.075}
\psline[linewidth=0.5pt,doubleline=true](-0.9,-0.2)(-0.9,0.2)
\psline[linewidth=0.5pt,doubleline=true](-0.1,-0.2)(-0.1,0.2)
\psline[linewidth=0.5pt,doubleline=true](1.4,-0.2)(1.4,0.2)
\psline[linewidth=0.5pt,doubleline=true](2.2,-0.2)(2.2,0.2)
\psline[linewidth=1.2pt]{*-*}(-1.2,0)(0.2,0)
\psline[linewidth=1.2pt]{*-*}(1,0)(2.5,0)
\end{pspicture}
  &  MBO, MBO
\\ \hline
\end{tabular}
\caption{Possible types of orbits in the Reissner-Nordstr\"{o}m-(anti-)de Sitter spacetime surrounded by a matter field with $\omega_q=-\frac{2}{3}$. The thick lines represent the range of $\tr$. $\tr=0$ is represented by a blank circle and the horizons by two vertical lines. Note that the cosmological horizon in types A to D is only present for $\tLambda>0$. The turning points are indicated by thick dots.}
\label{tab:orbit-types_066}
\end{center}
\end{table}
\begin{table}[H]
\begin{center}
\begin{tabular}{|lccll|}\hline
Type &  $0$'s & $\epsilon_{sign\left(\Lambda\right)}$   & Range of $\tr$ & Orbit \\
\hline\hline
 A  & 1 & $1_{\pm}$,$0_{\pm}$  &
\begin{pspicture}(-4,-0.2)(3.5,0.2)
\psline[linewidth=0.5pt]{-}(-4,0)(3.5,0)
\pscircle[hatchcolor=white,fillstyle=solid](-2,0){0.075}
\psline[linewidth=0.5pt,doubleline=true](-0.9,-0.2)(-0.9,0.2)
\psline[linewidth=0.5pt,doubleline=true](-0.1,-0.2)(-0.1,0.2)
\psline[linewidth=0.5pt,doubleline=true](2.2,-0.2)(2.2,0.2)
\psline[linewidth=1.2pt]{*-}(-1.2,0)(3.5,0)
\end{pspicture}
& TEO 
\\ \hline
B & 2 & $1_{-}$, $0_{-}$  &
\begin{pspicture}(-4,-0.2)(3.5,0.2)
\psline[linewidth=0.5pt]{-}(-4,0)(3.5,0)
\pscircle[hatchcolor=white,fillstyle=solid](-2,0){0.075}
\psline[linewidth=0.5pt,doubleline=true](-0.9,-0.2)(-0.9,0.2)
\psline[linewidth=0.5pt,doubleline=true](-0.1,-0.2)(-0.1,0.2)
\psline[linewidth=0.5pt,doubleline=true](2.2,-0.2)(2.2,0.2)
\psline[linewidth=1.2pt]{*-*}(-1.2,0)(0.2,0)
\end{pspicture}
& MBO
\\  \hline
 C & 3 & $1_{\pm}$, $0_{\pm}$   &
\begin{pspicture}(-4,-0.2)(3.5,0.2)
\psline[linewidth=0.5pt]{-}(-4,0)(3.5,0)
\pscircle[hatchcolor=white,fillstyle=solid](-2,0){0.075}
\psline[linewidth=0.5pt,doubleline=true](-0.9,-0.2)(-0.9,0.2)
\psline[linewidth=0.5pt,doubleline=true](-0.1,-0.2)(-0.1,0.2)
\psline[linewidth=0.5pt,doubleline=true](2.2,-0.2)(2.2,0.2)
\psline[linewidth=1.2pt]{*-*}(-1.2,0)(0.2,0)
\psline[linewidth=1.2pt]{*-}(1,0)(3.5,0)
\end{pspicture}
  & MBO, EO
\\ \hline
D & 4 & $1_{-}$  &
\begin{pspicture}(-4,-0.2)(3.5,0.2)
\psline[linewidth=0.5pt]{-}(-4,0)(3.5,0)
\pscircle[hatchcolor=white,fillstyle=solid](-2,0){0.075}
\psline[linewidth=0.5pt,doubleline=true](-0.9,-0.2)(-0.9,0.2)
\psline[linewidth=0.5pt,doubleline=true](-0.1,-0.2)(-0.1,0.2)
\psline[linewidth=0.5pt,doubleline=true](2.2,-0.2)(2.2,0.2)
\psline[linewidth=1.2pt]{*-*}(-1.2,0)(0.2,0)
\psline[linewidth=1.2pt]{*-*}(1,0)(1.8,0)
\end{pspicture}
  &  MBO, BO
\\ \hline
\end{tabular}
\caption{Possible types of orbits in the Reissner-Nordstr\"{o}m-(anti-)de Sitter spacetime surrounded by a matter field with $\omega_q=-\frac{1}{3}$. The thick lines represent the range of $\tr$. $\tr=0$ a is represented by blank circle and the horizons by two vertical lines. Note that the cosmological horizon is only present for $\tLambda>0$. The turning points are indicated by thick dots.}
\label{tab:orbit-types_033}
\end{center}
\end{table}

\section{Solution of the geodesic equation}\label{sec:solution}
In this section we present the analytical solution of the geodesic equation (Eq. \ref{6}) for each analyzed value of $\omega_q$. The right-hand side of Eq. (\ref{6}) is a polynomial of order $6$,
\begin{equation}
 P_{6}(\tr) = \sum_{i=1}^6 a_i\tr^i \,,
\label{eqn:poly}
\end{equation}
with the coefficients $a_i$ depending on the parameters of the black hole and on the value of $\omega_q$. The order of $P_6(\tr)$ can be reduced by $1$ with the substitution
\begin{equation}
 \tr = \pm\frac{1}{x}+\tr_{6} \,,
\label{eqn:substitution}
\end{equation}
where $\tr_6$ is a $0$ of $P_6(\tr)$. With this substitution Eq. \ref{6} becomes
\begin{equation}
 \left(x\frac{\dd x}{\dd \varphi}\right)^2=\sum_{i=0}^{5}b_i x^i = P_5(x) \,,
\label{eqn:dxdphi}
\end{equation}
with adjusted coefficients. By separating the variables, we can reformulate Eq. \ref{eqn:dxdphi} in terms of a hyperelliptic integral of the first kind,
 \begin{equation}
 \varphi-\varphi_{\rm in}=\int_{x_{\rm in}}^x{\frac{x'dx'}{\sqrt{P_5(x')}}} \,.
\label{eqn:int-r-equation}
\end{equation}
Eqn. \ref{eqn:int-r-equation} can be solved in terms of derivatives the Kleinian $\sigma$-function \cite{Hackmann:2008}
\begin{equation}
x = -\frac{\sigma_1(\vec{\varphi}_{\infty})}{\sigma_2(\vec{\varphi}_{\infty})}\,, 
\end{equation}
where $\sigma_i=\frac{\partial\sigma(\vec{v})}{\partial v_i}$ and
\begin{equation}
 \vec{\varphi}_{\infty} = \left(-\int_{x}^{\infty}{\frac{\dd x}{\sqrt{P_5(x)}}}, \varphi - \varphi_{\rm in}- \int_{x_{\rm in}}^{\infty}{{\frac{x \dd x}{\sqrt{P_5(x)}}}}\right)^T \, .
\end{equation}
By resubstituting $x$ into \ref{eqn:substitution}, we obtain the full solution
\begin{equation}
  \tr(\varphi)= \mp \frac{\sigma_2(\vec{\varphi}_{\infty})}{\sigma_1(\vec{\varphi}_{\infty})}+\tr_6 \,.
\label{eqn:solution}
\end{equation}
If we only look at orbits for light (e.g., $\epsilon=0$) the solution simplifies tremendously. In that case the right-hand side of Eq. (\ref{6}) reduces to a fourth order polynomial. With a similar substitution as before
\begin{equation}
  \tr = \pm\frac{1}{x}+\tr_{4} \,,
\end{equation}
where $\tr_{4}$ is a $0$ of $R(\tr)$, Eq. (\ref{6}) simplifies to
\begin{equation}
 \left(\frac{\dd x}{\dd \varphi}\right)^2=\sum_{i=0}^{4}c_ix^i\,.
\end{equation}
With a further substitution $x=\frac{1}{b_3}\left(4z-\frac{b_2}{3}\right)$ this reduces to the standard Weierstra{\ss} form
\begin{equation}
  \left(\frac{\dd z}{\dd \varphi}\right)^2=4z^3-g_2 z-g_3 \,.
\end{equation}
This equation can be solved in terms of the Weierstra{\ss} $\wp$-function \cite{Markushevich:1967}
\begin{equation}
 z(\varphi)=\wp(\varphi-\varphi';g_2,g_3)\,.
\end{equation}
Here $\varphi'=\varphi_{\text{in}}+\int_{z_{\text{in}}}^{\infty}{\frac{dz}{\sqrt{4z^3-g_2 z-g_3}}}$ only depends on the initial values $\varphi_{\text{in}}$ and $z_{\text{in}}$. A resubstitution leads to the full solution
\begin{equation}
 \tr=\pm\frac{c_3}{4\wp(\varphi-\varphi';g_2,g_3)-\frac{c_2}{2}}+\tr_4\,.
\end{equation}

\section{Orbits}
\begin{figure}[H]
\centering
	\subfigure[$\epsilon=1$, $\omega_q=-1$ $\ta=0.001$, $\tLambda=-0.07$, $\tQ=0.1$, $\tL=0.01$, $E=3.41$: Many-World Bound Orbit]{
		\includegraphics[width=0.41\textwidth]{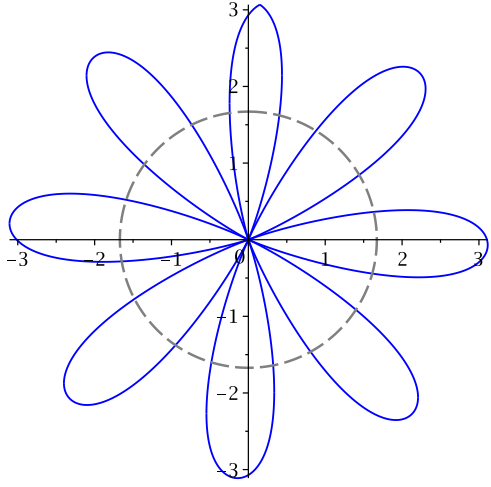}
	}
	\subfigure[$\epsilon=0$, $\omega_q=-\frac{2}{3}$, $\ta=0.03$, $\tLambda=-0.01$, $\tQ=0.1$, $\tL=0.04$, $E=\sqrt{0.93}$: Escape Orbit]{
		\includegraphics[width=0.41\textwidth]{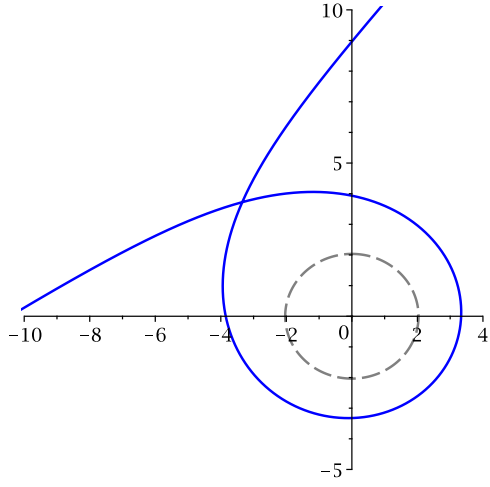}
	}

\subfigure[$\epsilon=1$, $\omega_q=-\frac{1}{3}$, $\ta=0.03$, $\tLambda=-0.01$, $\tQ=0.7$, $\tL=0.4$, $E=1.5$: Many-World Bound Orbit]{
		\includegraphics[width=0.41\textwidth]{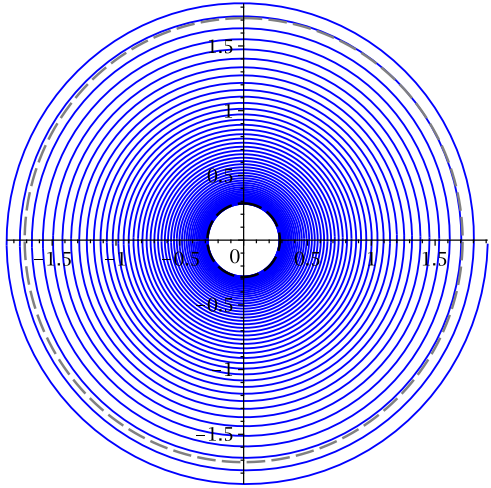}
	}
	\subfigure[$\epsilon=1$, $\omega_q=\frac{1}{3}$, $\ta=-0.03$, $\tLambda=-0.001$, $\tQ=0.1$, $\tL=0.04$, $E=1.1$: Bound Orbit]{
		\includegraphics[width=0.41\textwidth]{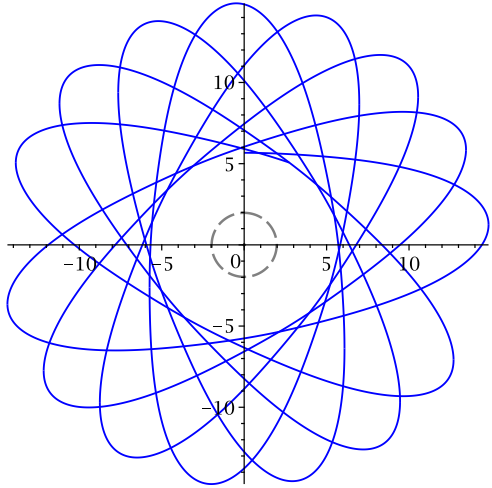}
	}

	\subfigure[$\epsilon=0$, $\omega_q=-\frac{2}{3}$, $\ta=0.1$, $\tLambda=-0.0025$, $\tQ=0.005$, $\tL=0.08$, $E=\sqrt{0.12}$: Two-World Escape Orbit]{
		\includegraphics[width=0.41\textwidth]{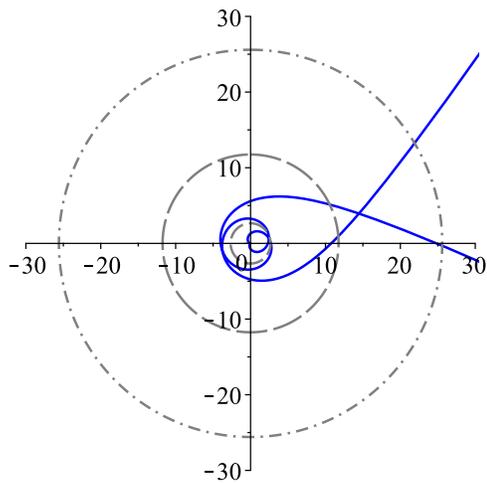}
	}
	\subfigure[$\epsilon=1$, $\omega_q=-\frac{2}{3}$, $\ta=0.1$, $\tLambda=-0.0025$, $\tQ=0.005$, $\tL=0.08$, $E=1.65$: Many-World Bound Orbit]{
		\includegraphics[width=0.41\textwidth]{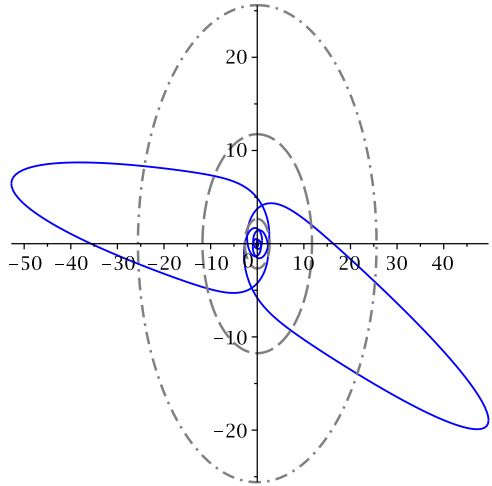}
	}
	\caption{Orbits of test particles and light in the Reissner-Nordstr\"{o}m-(anti-)de Sitter spacetime surrounded by various regular and exotic matter fields. The blue line corresponds to the geodesic and the dashed black lines denote the black hole horizons.}
 \label{pic:orbits}
\end{figure}

With the help of Eq. (\ref{eqn:solution}) we can visualize examples of possible orbits in the analyzed spacetime. Figure \ref{pic:orbits} shows various kinds of orbits for different values of the spacetime parameters.
\newpage
\section{Conclusion}
In this paper we studied the spacetime of a Reissner-Nordstr\"{o}m-(anti-)de Sitter black hole in the presence of different kinds of matter fields depending on different choices of the state parameter $\omega_q$ with the help of geodesics. After a short review of the structure of the spacetime, we derived the equation of motion for null and timelike geodesics for each separate matter fields. The analytical solutions were presented in terms of the elliptic Weierstra{\ss} $\wp$-function for null geodesics and in terms of derivatives of the Kleinian $\sigma$-function for timelike geodesics. Using effective potential techniques and parametric diagrams, a complete set of orbit types were analyzed for massive and massless test particles moving on geodesics. The derived orbits depend on the particle’s energy, angular momentum, cosmological constant and normalization parameter as well as on the state parameter. The analytical solutions obtained in Sec. \ref{sec:solution} are valid for all four values of $\omega_q$. The possible orbits for the spacetimes characterized by $\omega_q=\frac{1}{3}$ and $\omega_q=-1$ are extensively studied in \cite{Hackmann:2008tu} with the replacement of $Q^2$ by $Q^2-a$ and $\Lambda$ by $\Lambda+a$. In the view of geodesic motion taking place in the spacetime surrounded by irregular matter fields the possible orbit types are TEO, MBO, BO and EO for $\omega_q=-\frac{2}{3}$ and for $\omega_q=-\frac{1}{3}$. In the case of $\omega_q=-\frac{2}{3}$ it is possible to find MBO and TEO that crosses all four horizons. 
The results obtained in this paper are used to calculate the exact orbits and their properties. Furthermore, observables like periastron shift of bound orbits or the light deflection of escape orbits can be investigated. It would be interesting to extend the equation of motion and its solutions in the case of electrically and magnetically charged particles around the black hole considered in this paper. Especially orbits interacting with all four horizons should be investigated further. Another possibility for future work could be to study the influence of regular as well as exotic matter fields on the shadow cast by Reissner-Nordstr\"{o}m-(anti-)de Sitter black hole. We expect an influence of the state parameter $\omega_q$ and the normalization paramater $a$ on the the gravitational lensing of the black hole and the size of its shadow. Nevertheless these observables should be investigated in more detail in particular for the case of a positive cosmological constant in the future.

\section{Acknowledgements}
A.K.C and A.R. express sincere gratitude to  Dr. P. P. Pradhan who first made a strong impression on research related to the analytic study of black hole geodesic motion. K.F. thanks Jutta Kunz for interesting discussions and suggestions on the outline of the paper and gratefully acknowledges support by the DFG, within the research training group Models of Gravity. H.N. thanks the Science and Engineering Research Board (SERB), New Delhi for financial assistance during the course of this work through Grant No. EMR/2017/000339.

\bibliographystyle{unsrt}

\end{document}